\documentclass[twocolumn,pra,aps,superscriptaddress,10pt,longbibliography]{revtex4-2}
\usepackage{mathtools}
\usepackage{amsbsy}
\usepackage{amstext}
\usepackage[unicode=true,pdfusetitle,
 bookmarks=false,
 breaklinks=false,pdfborder={0 0 1},backref=false,colorlinks=false]
 {hyperref}
\usepackage[dvipsnames]{xcolor}
 \hypersetup{colorlinks=true,citecolor=RoyalPurple,filecolor=black,linkcolor=RoyalPurple,urlcolor=RoyalPurple}

\makeatletter

\@ifundefined{textcolor}{}
{%
 \definecolor{BLACK}{gray}{0}
 \definecolor{WHITE}{gray}{1}
 \definecolor{RED}{rgb}{1,0,0}
 \definecolor{GREEN}{rgb}{0,1,0}
 \definecolor{BLUE}{rgb}{0,0,1}
 \definecolor{CYAN}{cmyk}{1,0,0,0}
 \definecolor{MAGENTA}{cmyk}{0,1,0,0}
 \definecolor{YELLOW}{cmyk}{0,0,1,0}
}

\newcommand\ket[1]{\left|#1\right\rangle}
\newcommand\bra[1]{\left\langle #1 \right|}

\usepackage{amsmath}
\usepackage{graphicx}
\usepackage{amssymb}
\usepackage{txfonts,color}
\usepackage{dsfont}
\makeatother

\begin{document}
\title{Quantum switches for single-photon routing and entanglement generation\\ in waveguide-based networks}

\author{Juan Cumbrado}
\affiliation{Departamento de F{\'i}sica, Universidad Carlos III de Madrid, Avda. de la Universidad 30, 28911 Legan{\'e}s, Spain}

\author{Ricardo Puebla}
\affiliation{Departamento de F{\'i}sica, Universidad Carlos III de Madrid, Avda. de la Universidad 30, 28911 Legan{\'e}s, Spain}

\begin{abstract}
The interconnection of quantum nodes holds great promise for scaling up quantum computing units and enabling information processing across long-distance quantum registers. Such quantum networks can be realized using superconducting qubits linked by waveguides, which facilitate fast and robust on-demand quantum information exchange via traveling single photons. In this article, we propose leveraging additional qubit degrees of freedom as quantum switches that coherently condition the system dynamics. These switches are implemented using a qubit dispersively coupled to transfer resonators, which mediate interactions between node qubits and quantum links. Through wavepacket shaping techniques, we demonstrate that when the switch is closed, full excitation transfer occurs as a propagating photon, whereas an open switch allows only partial transfer without distorting the shape of the emitted photon. Based on this switch mechanism, we present deterministic protocols for generating entangled states via single-photon routing across the network, such as Bell, Greenberger-Horne-Zeilinger and W states. The feasibility of our approach is validated through numerical simulations of a three-node network, incorporating decoherence and photon loss effects. Our results indicate that high-fidelity entangled states can be realized employing the proposed quantum switches in current state-of-the-art platforms.
\end{abstract}

\maketitle

\section{Introduction}\label{s:intro}


The interconnection of nodes capable of hosting one or more quantum registers is a promising scalable architecture with applications across various fields~\cite{vanMeter}, including distributed quantum computing~\cite{Cirac1999,Caleffi2024}, quantum communications and the quantum internet~\cite{Kimble2008,Wehner2018}, quantum memories~\cite{Giovannetti2007,Reiserer2016}, and quantum metrology~\cite{Kwon2022,Morelli2022}. These interconnected nodes form a quantum network, which consist of spatially separated quantum registers connected by quantum links, enabling the exchange of quantum information. Recent experimental advancements have demonstrated the feasibility of such networks using trapped ions~\cite{Duan2010,Main2025}, atoms in cavities~\cite{Ritter2012,Nolleke2013,Reiserer2015}, superconducting qubits~\cite{Narla2016,Kurpiers2017,Axline2018,CampagneIbarcq2018,Zhong2019,Zhong2021,Storz2023,Niu2023,Qiu2025}, and color centers~\cite{Pfaff2014,Nguyen2019,Pompili2021}, all interconnected via emission and absorption of single bosonic quanta.


The control of information exchange and entanglement distribution across these quantum networks is an essential requisite to different applications, and thus finding primitive operations that are fast and noise resilient are highly valuable. In this context, we can distinguish between quantum teleportation-based protocols~\cite{Nolleke2013,Jiang2009,Hu2023} and those based on deterministic quantum state transfer~\cite{Cirac1996}, which have the potential of reducing the extra overhead introduced by qubit measurements. Quantum state transfer has been successfully achieved in networks made of superconducting qubits connected either via waveguides~\cite{Kurpiers2017,Magnard2020,Storz2023,Qiu2025} or through acoustic phonons~\cite{Bienfait2019,Chou2025} via wavepacket shaping protocols~\cite{Yin2013,Pechal2014,Penas2022,Penas2024b} where a single bosonic excitation with an engineered shape mediates the exchange of information among distant quantum registers.


In addition, designing quantum networks with additional elements can grant unique opportunities to enlarge the toolbox of primitive operations. A quantum switch is one of such elements. Albeit analogous to its classical counterpart, quantum switches allow for the coherent conditioning of the dynamical evolution of the system depending on their state. These switches can be therefore of relevance in routing flying qubits across the network and for entanglement distribution.  Previous works have put forward different schemes for the physical realization of a quantum switch, based either on dispersively coupled cavities~\cite{Wang2022}, interacting qubits to switch the transfer of a single photon from one to another transmission line~\cite{Rinaldi2025} or via tunable transmission-line resonators~\cite{Pechal2016}.

In this article, we propose the use of additional qubits dispersively coupled to transfer resonators that mediate the interconnection between nodes through quantum links. These additional qubits act as quantum switches, thus conditioning coherently the dynamical evolution of the system depending on their state (open or closed) as a consequence of the induced frequency shift on the transfer resonator. The scheme put forward here introduces a minimal modification to existing architectures and would allow to extend the primitive operations in these quantum networks beyond standard quantum state transfer protocols. 
As we show, the shape of the injected photon remains unaffected regardless of the switch state, a key requisite for entanglement distribution. Importantly, the amplitude of the injected photon for an open switch depends solely on the ratio between the frequency shift and the resonator decay rate or photon bandwidth. In this manner, quantum switches can be used for single-photon routing and thus entanglement generation across distant nodes of a quantum network. Based on the working principle of these quantum switches, we outline the requirements to engineer Bell, Greenberger-Horne-Zeilinger (GHZ) and W states. The feasibility of our approach is supported by numerical simulations of a three-node superconducting quantum network, inspired in recent experimental works such as Refs.~\cite{Storz2023,Qiu2025}, incorporating decoherence and photon loss effects. Our results indicate that high-fidelity entangled states can be realized employing the proposed quantum switches in current state-of-the-art platforms.


The article is organized as follows. In Sec.~\ref{s:qs} we present the working principle of a quantum switch and discuss its physical realization employing a qubit dispersively coupled to a transfer resonator under wavepacket shaping techniques. In Sec.~\ref{s:sp} we propose different protocols for entanglement generation, both bi- and multipartite, that leverage the conditional dynamics imposed by the quantum switches in the network, and single-photon routing. The suitability of the theoretical protocols to generate the targeted entangled states with high-fidelity is supported  by means of detailed numerical simulations in a three-node quantum network, including decoherence effects as $T_1$ qubit relaxation and photon loss. Finally, a summary of the main conclusions is presented in Sec.~\ref{s:conc}.

\section{Quantum switch via dispersive coupling}\label{s:qs}

\begin{figure}
    \centering
    \includegraphics[width=1\linewidth]{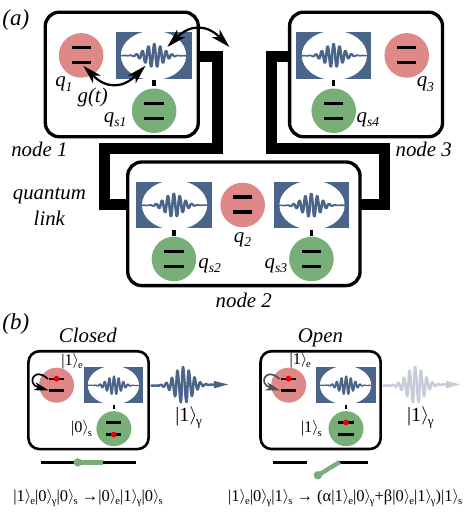}
    \caption{(a) Schematic illustration of a three-node quantum network consisting of node qubits $q_1$, $q_2$ and $q_3$ (red) coupled to transfer resonators (blue) that mediate the interaction between nodes across quantum links (thick black lines), through which single photons propagate. Each transfer resonator is coupled to an additional qubit degree of freedom that acts as a quantum switch (green). (b) Working principle of the emission protocol under the two situations of the quantum switch, either in its closed state $\ket{0}_s$ (left) or open state $\ket{1}_{s}$ (right). For the closed configuration the full excitation of the emitter or node qubit $\ket{1}_e$ is transformed as a propagating single photon through the quantum link (cf. Eq.~\eqref{eq:s1}). For the open switch, the excitation of the emitter is only partially injected as a single photon (cf. Eq.~\eqref{eq:s2}).  }
    \label{fig1}
\end{figure}

In this article, we focus on a superconducting implementation of a quantum network~\cite{Pechal2014,Kurpiers2017,Magnard2020,Storz2023,Qiu2025}, as illustrated in Fig.~\ref{fig1}(a) for a minimal linear network with three nodes.  There we can distinguish different elements. Quantum registers $q_j$ with $j\in\{1,2,3\}$ are connected to transfer resonators that in turn are coupled to their corresponding quantum link. These links are able to convey single photons across large distances, thus allowing for the exchange of quantum information on demand among spatially-separated registers. In addition, we also include an extra qubit degree of freedom $q_{s,j}$ per transfer resonator. These qubits act as quantum switches that condition the single-photon emission between nodes, thus opening the door for entanglement distribution protocols and single-photon routing across the network. To the contrary, closed switches at the receiver end ensure a full absorption of the incoming excitation. Therefore, one may relax the requirement of a quantum switch per transfer resonator if a particular direction of transmission is of interest, e.g. only from node 1 to node 2 but not vice versa. For completeness, however, we will discuss the role of all switches for the three-node network depicted in Fig.~\ref{fig1}(a). 

In particular, we refer to the action of a quantum switch as its ability to condition the single-photon emission in a coherent manner from an emitter in an excited state, say $\ket{1}_e$, into a propagating photon through the link, $\ket{1}_\gamma=\int d\omega f(\omega)b^\dagger_\omega \ket{{\rm vac}}$, see Fig.~\ref{fig1}(b). Here $b_\omega^\dagger$ refers to the creation bosonic operator of a mode of the quantum link with frequency $\omega$, while $f(\omega)$ is the photon frequency shape fulfilling $\int d\omega |f(\omega)|^2=1$, and $\ket{{\rm vac}}$ is the vacuum state. Ideally, an open switch should completely prevent the emission, while a closed switch should allow for the full photon transfer. However, the realization of a quantum switch that suppresses the emission may require stringent physical parameters. Moreover, an open quantum switch that partially prevents single photon transfer can be of interest to engineer relevant entangled states, as we show later. Without loss of generality, we refer to open when the switch qubit is in the excited state $\ket{1}_s$, so that $\ket{0}_s$ corresponds to the closed switch (cf. Fig.~\ref{fig1}(b)). Therefore, the emission operation conditioned to a partial quantum switch reads as
\begin{align}\label{eq:s1}
    \ket{1}_e \ket{0}_\gamma \ket{0}_s&\rightarrow \ket{0}_e\ket{1}_\gamma \ket{0}_s,\\ \label{eq:s2}
    \ket{1}_e \ket{0}_\gamma \ket{1}_s&\rightarrow(\alpha\ket{1}_e\ket{0}_\gamma+\beta\ket{0}_e\ket{1}_\gamma)\ket{1}_s,
\end{align}
where $|\beta|^2$ is the transmission probability for an open switch, with $|\alpha|^2+|\beta|^2=1$. 
In addition, in order to use the previous operation for entanglement distribution, the emitted photons in both cases must be indistinguishable.

A closed quantum switch allows for the full emission, and thus it can be used to implement standard quantum state transfer protocols upon the reabsorption of the propagating photon by a receiver qubit~\cite{Cirac1996}. Quantum state transfer refers to a unitary  $U_{qst}$ that exchanges the quantum information between qubits $i$ and $j$ in the following manner
\begin{align}\label{eq:qst}
    (\alpha\ket{0}_i+\beta\ket{1}_i) \ket{0}_j\xrightarrow[]{U_{qst}} \ket{0}_i(\alpha\ket{0}_j+\beta\ket{1}_j).
\end{align}
Typically, the realization of such transfer relies on a classical external variable, namely, either performing the set of protocols required to realize $U_{qst}$ or not. However, adding the quantum switch degree of freedom extends the possible protocols that can be realized, enlarging the capabilities for entanglement distribution and single-photon routing across the network, as we will show in Sec.~\ref{s:sp}.

The quantum switch is realized by a qubit dispersively coupled to the transfer resonator (cf. Fig.~\ref{fig1}(b)). For the sake of clarity, we briefly discuss the resulting Hamiltonian in the dispersive regime~\cite{Boissonneault2009}. The Hamiltonian of both interacting elements can be written as ($\hbar=1$)
\begin{align}\label{eq:Hqstr}
    H_{qs-tr}=\omega_{b-tr} a^\dagger a+\omega_{b-qs}\sigma_{qs}^+\sigma_{qs}^-+g(\sigma_{qs}^-a^\dagger+{\rm H.c.}),
\end{align}
where $\omega_{b-tr}$ and $\omega_{b-qs}$ denote the bare frequencies of the transfer resonator and quantum switch, respectively, while $g$ is the coupling strength. Here, the $a^\dagger$ and $a$ represent the creation and annihilation bosonic operators $[a,a^\dagger]=1$, while $\sigma_{qs}^+=\ket{1}\bra{0}$ is the raising spin operator for the quantum switch qubit. As customary in superconducting-based platforms, $g$ is much smaller than $\omega_{qs}$ and $\omega_{tr}$, so that counter-rotating terms in Eq.~\eqref{eq:Hqstr} can be safely neglected. In addition, we assume that both elements are in the dispersive regime, i.e. the detuning $\Delta=\omega_{b-qs}-\omega_{b-tr}$ is much larger than $g$, that is $|\Delta|\gg |g|$, so that $H_{qs-tr}$ can be well approximated by
\begin{align}
    H_{qs-tr}\approx \left(\omega_{b-tr}+\frac{g^2}{\Delta}\sigma^z_{qs} \right)a^\dagger a +\left(\omega_{b-qs}+\frac{g^2}{\Delta }\right)\frac{\sigma_{qs}^z}{2}+O\left(\frac{g^2}{\Delta^2}\right).\nonumber
\end{align}
The transfer resonator acquires a frequency shift dependent on the state of the quantum switch. Defining  $\chi=2g^2/\Delta$ as the total frequency variation on the resonator, we set its reference frequency to $\omega_{tr}=\omega_{b-tr}-\chi/2$. Without loss of generality, we assume the shift to be positive, $\chi\geq 0$. As the dispersive Hamiltonian is diagonal, the quantum switch becomes a passive element only conditioning the frequency of the transfer resonator according to $\omega_{tr}+\chi\sigma_{qs}^+\sigma_{qs}^-$. Such dispersive frequency shift plays a key role for readout of  superconducting qubits~\cite{Blais2021,Kono2017} or for realizing quantum gates~\cite{Penas2022}. 


\subsection{Wavepacking shaping}\label{ss:wv}
The next step consists in devising a control protocol that allows for the deterministic and fast exchange of an excitation from one node to another across the quantum link dependent on the state of the quantum switch. That is, an emission protocol that implements Eqs.~\eqref{eq:s1}-\eqref{eq:s2}. Although quantum state transfer has been achieved using adiabatic-like protocols~\cite{Baksic2016,Vogell2017,Leung19,Chang20} or by an always-on interaction~\cite{Serafini2006}, wavepacket shaping techniques allow for faster operation times with higher fidelities~\cite{Penas2022}. 
Wavepacket shaping techniques refer to the generation of a traveling photon with a desired shape via a suitable control pulse $g(t)$ that dictates the Jaynes-Cummings interaction between emitter and resonator~\cite{Ritter2012,Pechal2014,Zeytinoglu2015}. The shaped photon can then be reabsorbed at the receiver node by a time-reversed control pulse, following the seminal proposal in Ref.~\cite{Cirac1996}.  The emitter node consists of the emitter qubit coupled to the transfer resonator whose frequency depends on the state of the switch. For now we assume that the transfer resonator decays into the the quantum link via a Markovian decay rate $\kappa$, while numerical simulations beyond this simple description will be discussed in Sec.~\ref{s:num}. For a single excitation shared between resonant emitter qubit and transfer resonator, $\omega_{tr}=\omega_{e}$, the dynamics of the emitter node at the central frequency $\omega_{tr}$ follows from
\begin{align}\label{eq:qt}
    \dot{q}(t)&=-ig(t)c(t),\\\label{eq:ct}
    \dot{c}(t)&=-ig^*(t)q(t)-i\chi q_s c(t)-\kappa c(t)/2,
\end{align}
where $q(t)$ and $c(t)$ denote the amplitudes of the emitter and transfer resonator containing an excitation, respectively, while $q_s\in\{0,1\}$ is the state of the switch, $\ket{q_s}$. Note that the switch remains passive throughout the process. This simple model allows us to analytically derive the required control $g(t)$ given a photon shape either in the time or frequency domain, i.e. $\gamma(t)$ or $f(\omega)$, respectively.

In particular, for the closed switch case ($q_s=0$) and considering the initial conditions $\lim_{t\rightarrow-\infty}q(t)=1$ and $\lim_{t\rightarrow-\infty}c(t)=0$, i.e excited state and vacuum for the emitter and resonator, respectively,  the real control $g(t)=\kappa/2 \ {\rm sech}(\kappa t/2)$ generates a smooth sech-like photon with the maximum bandwidth $\kappa$, $|\gamma(t)|=\sqrt{\kappa/4}{\rm sech}(\kappa t/2)$ at the carrier frequency $\omega_{tr}$~\cite{Penas2022}. This follows from the standard input-output relation between the emitted photon and the resonator field, $\gamma(t)=-i\sqrt{\kappa}c(t)$~\cite{GardinerUltracoldII}. In this manner, the operation detailed in Eq.~\eqref{eq:s1} is realized, which enables quantum state transfer protocols employing a pitch-and-catch scheme, as experimentally demonstrated in previous works~\cite{Kurpiers2017,Storz2023}.

Importantly, for $q_s=1$ and $\chi\neq 0$, the emission process under the same $g(t)$ leads to a partial transfer of the excitation. We refer to Appendix~\ref{app:a} for the technical details of the derivation, and discuss here the main results. The probability of photon transmission $p_t$ depends solely on the ratio between  detuning and bandwidth $\chi/\kappa$, which reads as
\begin{align}\label{eq:pt}
    p_t=\frac{\kappa^2}{\chi^2+\kappa^2},
\end{align}
and the remaining probability in the emitter qubit is simply $p_r=1-p_t$. More specifically, the amplitudes $\alpha$ and $\beta$ in Eq.~\eqref{eq:s2} are given by $\alpha=(1-i(\kappa/\chi))^{-1}$, so that $p_r=|\alpha|^2$, and $\beta=(1+i(\chi/\kappa))^{-1}$, which fixes the transmission probability $p_t=|\beta|^2$ (cf. Eq.~\eqref{eq:pt}). As an example, for $\chi=\kappa$, the quantum switch in its open state ($q_s=1$) allows for just half of the excitation to go through as a traveling photon, $\alpha=e^{i\pi/4}/\sqrt{2}$ and $\beta=e^{-i\pi/4}/\sqrt{2}$, leading into a Bell state between emitter and photon (cf. Eq~\eqref{eq:s2}). We stress that the applied control does not depend on the quantum switch state, and it is always $g(t)=\kappa/2 {\rm sech}(\kappa t/2)$. Moreover, the working principle of the quantum switch as detailed in Eqs.~\eqref{eq:s1}-\eqref{eq:s2} demands indistinguishable photons regardless of the switch state, a condition that is fulfilled in this emission process. The emitted photon acquires the same shape independently of $q_s$ and $\chi$, while the coefficients $\alpha$ and $\beta$ can be controlled by tuning the ratio $\chi/\kappa$ (see Appendix~\ref{app:a} for more details). 

Before proceeding further, it is worth mentioning that the phenomenon explained in previous lines is not expected for a generic control $g(t)$. Indeed, adiabatic-like protocols~\cite{Leung19,Chang20} or those required to inject sech-like photons with a reduced bandwidth with respect to $\kappa$ will either feature a vanishing emitter excitation regardless the quantum switch state and/or produce distinguishable photons dependent on $\chi$ (see Appendix~\ref{app:a}).

\section{Entanglement generation and single-photon routing}\label{s:sp}
The operation of the quantum switch under the emission protocol discussed previously extends the possibilities for single-photon routing and entanglement generation across the network. Here we detail the protocols required for the generation of three important classes of states, namely, Bell states in Sec.~\ref{ss:Bell}, and GHZ and W states in Sec.~\ref{ss:multi}. In addition, in Sec.~\ref{ss:sp_routing} we also discuss how to route a single excitation based on the states of the quantum switches. The suitability of the proposed protocols to reach competitive fidelities in state-of-the art setups is supported by means of detailed numerical simulations, including realistic imperfections and decoherence effects, see Sec.~\ref{s:num}.

\subsection{Bell states}\label{ss:Bell}
Bipartite entangled states between two qubits on different nodes can be achieved in a deterministic manner as follows. First, the emitter qubit and receiver qubits are initialized in $\ket{1}_1$ and $\ket{0}_2$, respectively. Second, the quantum switch of the emitter is open, $\ket{1}_{s_1}$, so that upon the emission protocol $g(t)$, it only allows for a fraction of the excitation to propagate through the quantum link. Third, the incoming photon is then reabsorbed by the receiver node by a delayed and time-reversed emission protocol $g(t_p-t)$, where $t_p$ accounts for the propagation time of the photon between nodes. The receiver node, with a closed quantum switch $\ket{0}_{s_2}$, is then able to fully absorb the incoming excitation. The emission process is described by Eq.~\eqref{eq:s2}, which is followed by the subsequent reabsorption resulting in
\begin{align}
&\ket{1}_1\ket{0}_2\ket{1}_{s_1}\ket{0}_{s_2}\nonumber\\&\rightarrow \left(\frac{\chi}{\chi-i\kappa}\ket{1}_1\ket{0}_2+\frac{e^{i\phi}\kappa}{\kappa-i\chi}\ket{0}_1\ket{1}_2\right)\ket{1}_{s_1}\ket{0}_{s_2}.
\end{align}
Note that we have made used of the explicit expressions for the coefficients $\alpha$ and $\beta$ described previously, while $e^{i\phi}$ represents the phase acquired by the traveling photon during the dynamical process. Therefore, when the switch at the emitter node induces a dispersive shift equal to the photon bandwidth,  $\chi=\kappa$, the previous expression reduces to a Bell state,
\begin{align}
\ket{1}_1\ket{0}_2\ket{1}_{s_1}\ket{0}_{s_2}\rightarrow \frac{1}{\sqrt{2}}\left(\ket{1}_1\ket{0}_2+e^{i\phi'}\ket{0}_1\ket{1}_2 \right)\ket{1}_{s_1}\ket{0}_{s_2},\nonumber
\end{align}
with $\phi'=\phi-\pi/2$. We note that, although deterministic generation of Bell states can be achieved without the need for extra degrees of freedom~\cite{Kurpiers2017,CampagneIbarcq2018,Bienfait2019,Zhong2019,Zhong2021,Storz2023,Penas2024}, quantum switches allow for the simultaneous realization of standard quantum state transfer \textit{and} bipartite entanglement generation. This can be easily shown since the external controls $g(t)$ are the same regardless of the switch at the emitter node, thus naturally unlocking the engineering of multipartite entangled states.

\subsection{Multipartite entanglement}\label{ss:multi}

The previous protocol can be generalized when the switch at the emitter node is in a generic state $\ket{\psi}_{s_1}=\alpha_s\ket{0}_{s_1}+\beta_s\ket{0}_{s_1}$, so that $|\alpha_s|^2+|\beta_s|^2=1$. Following the same steps as in Sec.~\ref{ss:Bell}, the initial state $\ket{1}_{1}\ket{0}_2\ket{\psi}_{s_1}\ket{0}_{s_2}$ is transformed into $\ket{\Psi}$, which reads as
\begin{align}
    \ket{\Psi}=&\left[\alpha_s e^{i\phi}\ket{0}_1\ket{1}_2\ket{0}_{s_1}+ \right.\nonumber \\& \left. \beta_s\left(\frac{\chi}{\chi-i\kappa}\ket{1}_1\ket{0}_2 +\frac{e^{i\phi}\kappa}{\kappa-i\chi}\ket{0}_1\ket{1}_2\right)\ket{1}_{s_1} \right]\ket{0}_{s_2}.
\end{align}
This state contains genuine multipartite entanglement depending on the specific parameters $\alpha_s,\beta_s,\chi$ and $\kappa$. In particular, a GHZ state can be approximately realized when $\chi/\kappa\gg 1$. In this limit, the previous state $\ket{\Psi}$ can be approximated as
\begin{align}
    \ket{\Psi}\approx \left[\alpha_s e^{i\phi}\ket{0}_1\ket{1}_2\ket{0}_{s_1}+\beta_s\ket{1}_1\ket{0}_2\ket{1}_{s_1}\right]\ket{0}_{s_2},
\end{align}
which upon a local $X_2$ gate (on the second qubit) and for $\ket{\psi}_{s_1}=\frac{1}{\sqrt{2}}(e^{-i\phi}\ket{0}_{s_1}+\ket{1}_{s_1})$, it corresponds to the three-qubit GHZ state $|\Psi_X\rangle=X_2\ket{\Psi}\approx \ket{{\rm GHZ}_3}=\frac{1}{\sqrt{2}}(\ket{000}+\ket{111})$, between the node qubits 1 and 2 and the quantum switch at the emitter node. Yet, unlike Bell states, the generation of a $\ket{{\rm GHZ}_3}$ state is only achieved approximately depending on the largest achievable ratio $\chi/\kappa$. The fidelity of the generated state with respect to the target $\ket{{\rm GHZ}_3}$ state results in $F=|\langle {\rm GHZ}_3|\Psi_X\rangle|^2= (1+3\chi^2/(\chi^2+\kappa^2))/4$, which tends to $1$ as $\chi/\kappa\rightarrow \infty$. This fidelity can be further improved, allowing for a relative phase in the GHZ state, or alternatively, performing a local-phase gate $P=\ket{0}\bra{0}+e^{i\varphi}\ket{1}\bra{1}$ in the final state to compensate the phase $\varphi$ in the $\alpha$ coefficient, i.e. $\alpha=|\alpha|e^{-i\varphi}$. Taking into account this phase correction,  $|\tilde{\Psi}\rangle=P_2X_2\ket{\Psi}$, the fidelity with respect to a targeted GHZ state reads as
\begin{align}\label{eq:FGHZ}
    F=|\langle {\rm GHZ}_3|\tilde{\Psi}\rangle|^2=\frac{1}{4}\left(1+\left(\frac{\chi^2}{\chi^2+\kappa^2}\right)^{1/2} \right)^2.
\end{align}
As the ratio between the dispersive shift and the decay rate $\kappa$ grows, the fidelity tends to $1$, $F\approx 1-\kappa^2/(2\chi^2)+O(\kappa^4/\chi^4)$ for $\chi/\kappa\gg 1$. Note that already for $\chi>\sqrt{ (3-2\sqrt{2})/(2\sqrt{2}-2)}\kappa\approx 0.455\kappa$ one would generate states with genuine GHZ entanglement, as revealed by the fidelity witness $F>1/2$ upon phase correction~\cite{Guhne2009}.



We now move to the engineering of another relevant multipartite entangled state, namely, the $\ket{W_N}$ state among $N$ qubits. This state can be written as $\ket{W_N}=\frac{1}{\sqrt{N}}(\ket{00\ldots 1}+\ldots+\ket{01\ldots 0}+\ket{10\ldots 0})$. In this case, multipartite entanglement can be generated among the node qubits, by a suitable choice of the dispersive frequency shifts induced by the switches and using a sequential emission-absorption protocol between neighboring nodes of a linear network. For simplicity, we assume equal decay rates along the network. In particular, an initial state $\ket{1}_1\ket{0}_2\ldots\ket{0}_N$ can be transformed into the $\ket{W_N}$ state (up to local phases) by means of $N-1$ emission-absorption protocols. The quantum switches of the receivers must be closed to fully absorb the incoming photon, i.e. $\ket{0}_{s_{2k}}$ for $k=1,\ldots, N-1$, while the $(2k-1)$th switch at the emitting node is open and allows for an excitation injection with probability
\begin{align}
    p_t^{(2k-1)}=\frac{N-k}{N+1-k} \quad k=1,\ldots, N-1.
\end{align}
This requirement directly translates into a condition for the dispersive shift (cf. Eq.~\eqref{eq:pt}),
\begin{align}
    \chi^{(2k-1)}=\frac{\kappa}{\sqrt{N-k}}\quad k=1,\ldots, N-1,
\end{align}
while $\chi^{(2k)}$ is irrelevant since the receiver switch is closed.  
In particular, for the preparation of a $\ket{W_3}$ state among $N=3$ qubits, the first quantum switch must be such that $\chi^{(1)}=\kappa/\sqrt{2}$, while the third switch $\chi^{(3)}=\kappa$ for the subsequent emission. In this process, the quantum switches do not become entangled and remain in their initial state, i.e. $\ket{1}_{s_1}\ket{0}_{s_2}\ket{1}_{s_3}\ket{0}_{s_4}$. 

\subsection{Directional single-photon routing}\label{ss:sp_routing}

Quantum switches can be of relevance for the directionally routing of a single photon.  For that we focus on the central node in Fig.~\ref{fig1}(a), which can perform a left and right emission. As before, we assume a simple Markovian decay model for this central node at the resonant frequency $\omega_{tr}$. To ease the notation, we refer to left and right transfer resonators, controls and switches by subscripts $l$ and $r$, respectively. The dynamics is described by
\begin{align}
    \dot{q}(t)&=-ig_l(t)c_l(t)-ig_r(t)c_r(t)\\
    \dot{c}_l(t)&=-ig_l(t)q(t)-i q_{s_l} \chi_{s_l} c_l(t)-\kappa c_l(t)/2\\
    \dot{c}_r(t)&=-ig_r(t)q(t)-i q_{s_r} \chi_{s_r} c_r(t)-\kappa c_r(t)/2,
\end{align}
where we have already assumed equal decay rates $\kappa$ and a real controls $g_l(t),g_r(t)\in\mathbb{R}$. The initial excitation of the emitter can be transformed into two photons with equal amplitude propagating in opposite directions under a global and simultaneous control $g_r(t)=g_l(t)=\kappa/(2\sqrt{2}){\rm sech}(\kappa t/2)$~\cite{Penas2024} for both closed switches, $q_{s_l}=q_{s_r}=0$. In addition, if both dispersive shifts are equal, $\chi_{s_l}=\chi_{s_r}=\chi$, one recovers a similar situation as outlined in Eqs.~\eqref{eq:s1}-\eqref{eq:s2} provided the switches are both closed or both open. Unfortunately, a simultaneous protocol fails for different switch states as it no longer guarantees the emission of indistinguishable photons. To circumvent this issue, the protocol must be done sequentially using again $g_{l,r}(t)=\kappa/2 \ {\rm sech}(\kappa t/2)$, i.e. first the left and then right, or vice versa. For a left-right sequence (the reverse scenario can be obtained by simply swapping the left-right subscripts), one finds the following operation
\begin{widetext}
\begin{align}
\ket{1}_e\ket{0}_{\gamma_l}\ket{0}_{\gamma_r}\ket{0}_{s_l}|q\rangle_{s_r}&\rightarrow \ket{0}_e\ket{1}_{\gamma_l}\ket{0}_{\gamma_r}\ket{0}_{s_l}|q\rangle_{s_r},\quad q\in\{0,1\}\\
\ket{1}_e\ket{0}_{\gamma_l}\ket{0}_{\gamma_r}\ket{1}_{s_l}|0\rangle_{s_r}&\rightarrow (\alpha_l \ket{0}_e\ket{0}_{\gamma_l}\ket{1}_{\gamma_r}+\beta_l\ket{0}_e\ket{1}_{\gamma_l}\ket{0}_{\gamma_r})\ket{1}_{s_l}|0\rangle_{s_r}\\
\ket{1}_e\ket{0}_{\gamma_l}\ket{0}_{\gamma_r}\ket{1}_{s_l}|1\rangle_{s_r}&\rightarrow (\alpha_l \alpha_r\ket{1}_e\ket{0}_{\gamma_l}\ket{0}_{\gamma_r}+\alpha_l\beta_r \ket{0}_e\ket{0}_{\gamma_l}\ket{1}_{\gamma_r}+\beta_l\ket{0}_e\ket{1}_{\gamma_l}\ket{0}_{\gamma_r})\ket{1}_{s_l}|1\rangle_{s_r},
\end{align}
\end{widetext}
where the expressions for $\alpha_{l,r}$ and $\beta_{l,r}$ are equivalent to those of a single quantum switch case but with the specific dispersive shift of each switch $\chi_{l,r}$ (cf. Sec.~\ref{ss:wv}).  In addition, note that since the protocol is done sequentially, the controls are delayed (assuming no overlap) leading into delayed left-right traveling photons, as well as with the corresponding phases due to the dynamics. Analogous to the previous discussion on GHZ state engineering, depending on the ratio $\chi_{l,r}/\kappa$ the emission can be largely suppressed in one direction for an open switch.

\section{Numerical simulations}\label{s:num}
In the following we benchmark the suitability of the proposed protocols to generate entanglement leveraging the quantum switches by numerically simulating a quantum network. For that we first introduce the full model in Sec.~\ref{ss:model}. Decoherence effects are discussed in Sec.~\ref{ss:opt}, where we also introduce an optimal operation time, which is then used for further numerical simulations presented in Sec.~\ref{ss:ent_num} that showcase the high fidelities with which two- and three-qubit entangled states can be generated.

\subsection{Model}\label{ss:model}
The setup, as illustrated in Fig.~\ref{fig1}(a), is based on current quantum networks experiments involving superconducting qubits operating at a temperature of few mK~\cite{Kurpiers2017,Magnard2020,Storz2023}. In particular, the Hamiltonian of a three-node linear quantum network can be written as
\begin{align}
    H=\sum_{j=1}^3 H_{n_j}(t)+\sum_{j=1}^2 \left[H_{QL_j}+H_{n-QL_j}\right],
\end{align}
where $H_{n_j}(t)$ describes the $j$th local node Hamiltonian, $H_{QL_j}$ the $j$th quantum link or superconducting waveguide that mediates the single microwave photon exchange, and $H_{n-QL_j}$ refers to the interaction between nodes and quantum link. In particular, at the resonant frequency $\omega_{tr}$ of the qubits and transfer resonators, the time-dependent Hamiltonian of first and third nodes is given by 
\begin{align}
    H_{n_1}(t)&=\omega_{tr}\sigma_1^+\sigma_1^-+(\omega_{tr}+\chi_{s_1}q_{s_1})a^\dagger_1 a_1+(g_1(t)\sigma^+_1a_1+{\rm H.c.}),\nonumber\\
    H_{n_3}(t)&=\omega_{tr}\sigma_3^+\sigma_3^-+(\omega_{tr}+\chi_{s_4}q_{s_4})a^\dagger_4 a_4+(g_4(t)\sigma^+_3a_4+{\rm H.c.}),\nonumber
\end{align}
while the Hamiltonian of the central node reads
\begin{align}
    H_{n_2}(t)=\omega_{tr}\sigma_2^+\sigma_2^-&+(\omega_{tr}+\chi_{s_2}q_{s_2})a_2^\dagger a_2+(\omega_{tr}+\chi_{s_3}q_{s_3})a_3^\dagger a_3\nonumber\\&+(g_2(t)\sigma_2^+a_2+g_3(t)\sigma_2^+ a_3+{\rm H.c.}).\nonumber
\end{align}
Here the coefficients $q_{s_j}\in\{0,1\}$ correspond to the state of the $j$th quantum switch which remains passive during the protocol  (cf. Sec.~\ref{s:qs}), while $\sigma_j^+$ and $a_j^\dagger$ is the spin raising and bosonic creation operator of the $j$th  qubit and transfer resonator, respectively. The quantum link is represented by two equal WR90 superconducting waveguides~\cite{Kurpiers2017,Magnard2020,Storz2023}, so that
\begin{align}
    H_{QL_j}=\sum_{k=1}^{N_m}\omega_k b^{(j),\dagger}_k b^{(j)}_k,
\end{align}
being $\omega_k=c_{light}\sqrt{(\pi/l_c)+(k\pi/L)^2}$ the frequency  of the ${\rm TEM}_{10k}$ mode in the WR90 waveguide with broad wall dimension $l_c=0.0286$ m and total length $L$, $c_{light}$ denoting the speed of light in vacuum~\cite{Pozar,Magnard2020}. Each mode has its corresponding bosonic operators for the $j=1,2$ quantum link, obeying $[b_i^{(k)},b_j^{(l),\dagger}]=\delta_{i,j}\delta_{k,l}$. The interaction between quantum link and nodes reads as
\begin{align}
    H_{n-QL_1}&=\sum_{k=1}^{N_m}G_k\left[a_1^\dagger b_k^{(1)}+(-1)^ka_2^\dagger b_k^{(1)}+{\rm H.c.}\right]\\
    H_{n-QL_2}&=\sum_{k=1}^{N_m}G_k\left[a_3^\dagger b_k^{(2)}+(-1)^ka_4^\dagger b_k^{(2)}+{\rm H.c.}\right],
\end{align}
where $G_k=\sqrt{\kappa v_g \omega_k/(2\omega_{tr}L)}$ is the coupling strength between the $k$th waveguide mode and the transfer resonators, assumed to decay at rate $\kappa$, while $v_g$ is the group velocity at the carrier frequency.  For a carrier frequency of $\omega_{tr}=2\pi\times 8$ GHz, it amounts to $v_g\approx 2c_{light}/3$, so that the time for a microwave photon to propagate between nodes separated by $L=10$ m is $t_p=L/v_g\approx 50$ ns. The alternating sign in the coupling gives account for the parity of the ${\rm TEM}_{10k}$ modes at the end points, while $N_m$ refers to the number of modes included in the simulation to ensure numerical convergence.  For a $L=10$ m long waveguide, the free spectral range amounts to $\Delta_{fsr}\approx 2\pi\times 8.6$ MHz at the carrier frequency $\omega_{tr}=2\pi\times 8$ GHz, and we choose $\kappa=2\pi\times 10$ MHz, close to experimental parameters~\cite{Kurpiers2017,Magnard2020,Storz2023}, although we stress that similar results can be found for other lengths, frequencies and decay rates.  Note that these physical parameters justify the absence of counter-rotating interaction terms. 

The dynamics of the full quantum network is solved in the single-excitation subspace using the Wigner-Weisskopf Ansatz, 
\begin{align}\label{eq:WW}
\ket{\Psi(t)}=&\left[\sum_{i=1,3}q_i(t)\sigma_i^++\sum_{i=1,4}c_i(t)a_i^\dagger+\right.\nonumber\\&\left.+\sum_{j=1}^2\sum_{k=1}^{N_m}\psi_k^{(j)}(t)b_k^{(j),\dagger}\right]\ket{{\rm vac}}\otimes_{j=1}^4|q_{s_j}\rangle_{s_j},
\end{align}
where $q_i(t)$, $c_i(t)$, and $\psi_k^{(j)}(t)$ refer to the amplitudes of a single excitation in the $i$th qubit, transfer resonator or $k$th mode of the $j$th waveguide, respectively, ensuring a normalized state, $\langle\Psi(t)|\Psi(t)\rangle=1$, i.e. $\sum_{i,k,j} (|q_i(t)|^2+|c_i(t)|^2+|\psi_k^{(j)}|^2)=1$. In addition, $\ket{{\rm vac}}$ denotes the vacuum state for all the elements, besides the quantum switches, initialized in states $|q_{s_j}\rangle_{s_j}$ depending on the protocol. The dynamics of these amplitudes $q_i(t),c_i(t),\psi_k(t)\in\mathbb{C}$ follows from $\frac{d}{dt}\ket{\Psi(t)}=-iH\ket{\Psi(t)}$, resulting in a set of linear coupled differential equations. As discussed in Secs.~\ref{s:qs} and~\ref{s:sp}, the dynamics of these amplitudes depends on the specific states of the quantum switches. 

\begin{figure}
    \centering
    \includegraphics[width=1\linewidth]{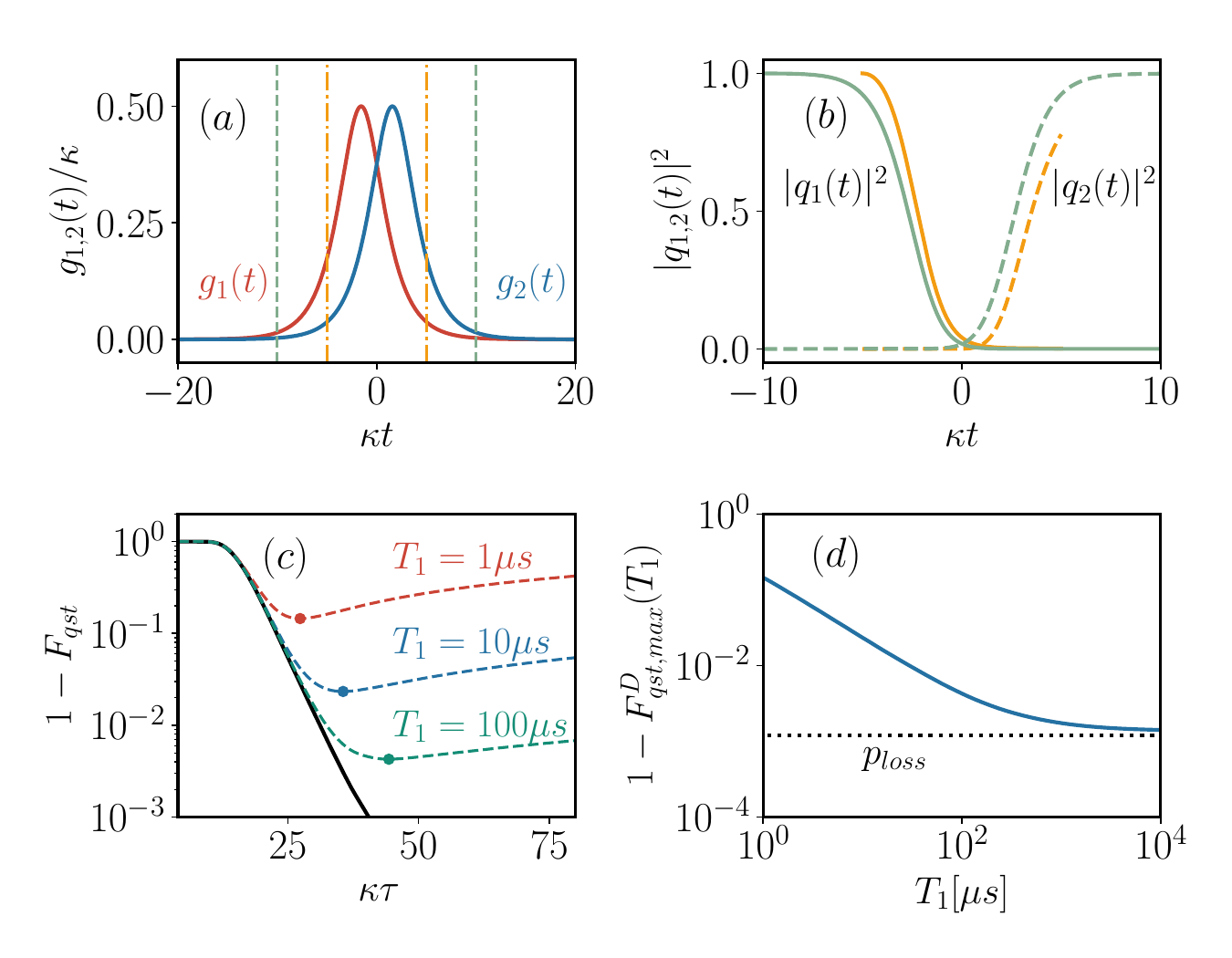}
    \caption{Panel (a) shows the profile of the control pulses $g_{1,2}(t)=\kappa/2 \ {\rm sech}(\kappa (t\pm t_p)/2)$ for a waveguide of $L=10$ m and $\kappa=2\pi\times 10$ MHz, delayed by the propagation time, $\kappa t_p\approx 3.14$.  Vertical lines indicate the cut-off at $\pm \kappa \tau/2$ for two different protocol times $\kappa\tau=20$ (green dashed) and $10$ (orange dot-dashed), which lead into the state transfer dynamics for the qubit populations depicted in (b), that spans the range $t\in[-\tau/2,\tau/2]$. In (c) the infidelity for the coherent state transfer $1-F_{qst}(\tau)$ (solid black line) is compared to the infidelities for $p_{loss}=1.2\cdot 10^{-3}$ and three different $T_1$ times, $1-F_{qst}^D(T_1,\tau)$. The solid points mark the position of the optimal operation time $\tau_{\rm opt}$ at which the fidelity is maximized. The minimum infidelity for a given $T_1$, $1-F_{qst,max}(T_1)$, is shown in panel (d), which is limited by photon loss $p_{loss}$ (dotted line).  See main text for further details.}
    \label{fig2}
\end{figure}

\subsection{Decoherence and optimal operation time}\label{ss:opt}
The main sources of errors in this setup stem from $T_1$ coherence time of the qubits, as well as from photon losses in the waveguide. Photon loss has been measured in recent experiments, yielding an estimated attenuation in the range of $0.3 -1$ dB/km~\cite{Qiu2025,Storz2023}. For our simulations we take a realistic and fixed value of $0.5$ dB/km, or equivalently $p_{\rm loss}=0.12\%$ of photon loss for a $L=10$ m long waveguide. To the contrary, we investigate the dependence of the overall fidelities on $T_1$ coherence times, assuming $T_2$ to be $T_1$-limited. 

Ideally, wavepacket shaping protocols require a total protocol time $\tau$ longer than $1/\kappa$ to reliably emit the desired shaped photon, and also to be able to fully absorb the incoming excitation. Yet, faster protocols are preferred whenever a finite $T_1$ time is considered. This natural trade-off between protocol and coherence time, $\tau$ vs $T_1$, leads into an optimal operation time $\tau_{\rm opt}$. Since the quantum switches allow for distinct operations across the network, $\tau_{\rm opt}$ might differ depending on the specific targeted state. However, as a fair optimization, we obtain $\tau_{\rm opt}$ focusing on a standard quantum state transfer between two nodes (cf. Eq.~\eqref{eq:qst}). Such operation time $\tau_{\rm opt}$ will be later employed for other protocols where it might not be exactly optimal.

To find the optimal protocol time $\tau_{\rm opt}$ we perform numerical simulations of a standard quantum state transfer varying the protocol time $\tau$.  An initial excitation in qubit 1, $q_1(-\tau/2)=1$, is mapped onto the initially empty state of the qubit 2, $q_2(-\tau/2)=0$, only applying controls on $g_1(t)$ and $g_2(t)$ with closed switches, $q_{s_1}=q_{s_2}=0$. For the emission-absorption controls we apply $g_1(t)=\kappa/2 \ {\rm sech}(\kappa (t+t_p/2)/2)$ and $g_2(t)=\kappa/2 \ {\rm sech}(\kappa(t-t_p/2)/2)$ for $t\in[-\tau/2,\tau/2]$, which are delayed by the propagation time $t_p$ (cf. Fig.~\ref{fig2}(a)).

The fidelity of the coherent state transfer is simply $F_{qst}(\tau)=|q_2(\tau/2)|^2$, i.e. given by the population of the receiver qubit at the end of the protocol. The effect of a finite $T_1$ coherence time and photon loss spoils the fidelity according to
\begin{align}\label{eq:Fqst}
    F_{qst}^D(T_1,\tau)=(1-p_{\rm loss})F_{qst}(\tau)e^{-(p_1(\tau)+p_2(\tau))/T_1},
\end{align}
where $p_j(\tau)=\int_{-\tau/2}^{\tau/2}dt|q_j(t)|^2$ is the time-integrated population of the $j$th qubit during the protocol. This simple expression stems from the fact that when either an incoherent transition of the qubits takes place $\ket{1}_{1,2}\rightarrow \ket{0}_{1,2}$ or the photon is lost, the system is brought to the vacuum state. 
In Fig.~\ref{fig2}(a) we show the controls $g_{1,2}(t)$ indicating the cut-off imposed by a finite duration $\tau$ for two cases, $\kappa\tau=10$ and $20$, while Fig.~\ref{fig2}(b) illustrates their impact on the dynamics of qubit populations $|q_{1,2}(t)|^2$ in a standard state transfer protocol. The infidelity of the coherent state transfer $1-F_{qst}(\tau)$ is compared to resulting infidelity with photon loss and finite $T_1$ time, $1-F_{qst}^D(T_1,\tau)$, in Fig.~\ref{fig2}(c). The maximum fidelity for a given $T_1$, $F_{qst,max}^D(T_1)=\max_\tau F_{qst}^D(T_1,\tau)$, marks the position of the optimal operation time, $\tau_{\rm opt}={\rm argmax}_\tau F^D_{qst}(T_1,\tau)$. For the considered parameters, this results in $\tau_{\rm opt}\approx 31\log(T_1)$ ns for $T_1\in[10^0,10^4]\ \mu$s, e.g. $\tau_{\rm opt}\approx 357$ ns  for $T_1=100\ \mu$s. The minimum infidelity $1-F_{qst,max}^D(T_1)$  is shown in Fig.~\ref{fig2}(d) as a function of $T_1$. Note that, for coherence times $T_1\gtrsim 400\ \mu$s, the fidelity is limited by photon loss, whose effect is more relevant in this parameter regime than photon distortion produced by the non-linear dispersion relation of the waveguide~\cite{Penas2022,Penas2023}. Indeed, photon distortion results in a minimum coherent infidelity, $1-F_{qst,max}\approx 10^{-7}$, which is much smaller than $p_{loss}$. The optimal time will be used for the following numerical simulations aimed at generating different entangled states.

\subsection{Quantum switch-based protocols}\label{ss:ent_num}

\begin{figure*}
    \centering
    \includegraphics[width=1\linewidth]{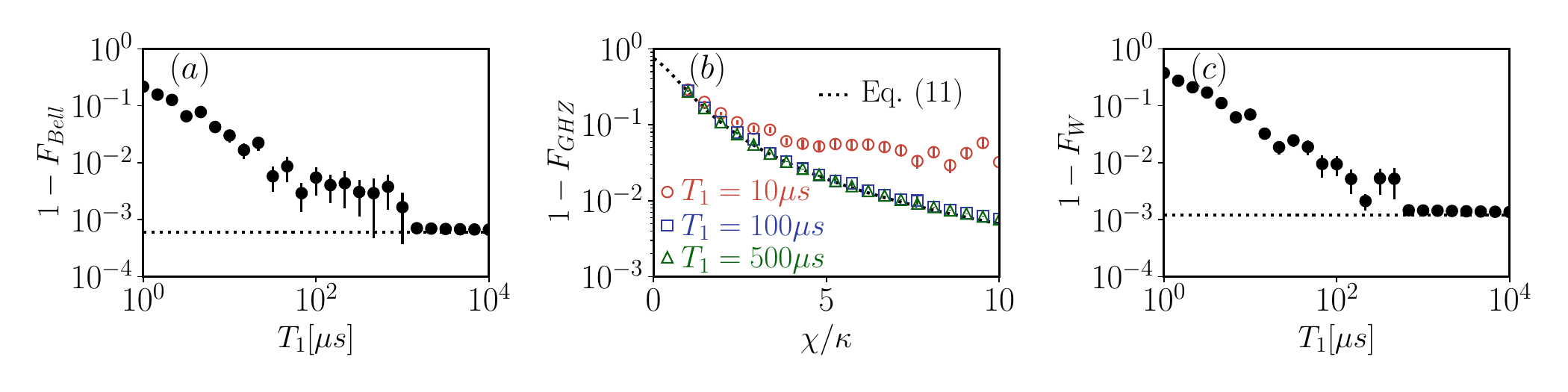}
    \caption{Numerical results using bootstrapping to estimate fidelities for different targeted entangled states using $N=1000$ quantum trajectories. Error bars correspond to a standard deviation obtained from $R=100$ repetitions of $M=500$ randomly chosen samples.  Panel (a) shows the infidelity $1-F_{\rm Bell}$ to generate a Bell state between first and second qubits as a function of the $T_1$ noise affecting all qubits in the network, as well as photon loss, which limits the fidelity to $1-F_{Bell}\approx p_{loss}/2$ (dotted line). The resulting fidelity for $T_1\gtrsim 10^2\ \mu$s is already above $99\%$. Panel (b) shows the infidelity targeting a $\ket{{\rm GHZ}_3}$ state, between first switch qubit and first and second node qubits, as a function of the ratio $\chi/\kappa$ and for three different $T_1$ times. The dotted line corresponds to the theoretical expression for the fidelity given in Eq.~\eqref{eq:FGHZ}. All the points are above the fidelity witness $F>1/2$ revealing genuine GHZ entanglement. Panel (c) corresponds to a target $\ket{W_3}$ state between the three node qubits, whose fidelity is limited by photon loss (dotted line). As for Bell states, for $T_1\gtrsim 10^2 \ \mu$s we find $F_{W}\gtrsim 0.99$. The protocol duration is set to $\tau=\tau_{\rm opt}$ (i.e. at the optimal operation time) for (a) and (b), while in (c) it is extended to $\tau=2\tau_{\rm opt}$ to take into account its sequential nature. See main text for further details. }
    \label{fig3}
\end{figure*}

We numerically test the performance of the proposed quantum switch-based protocols to generate Bell, GHZ and W states as a function of $T_1$ relaxation and photon loss by computing the resulting fidelity of the relevant state with respect to the target one. Yet, as the dynamics for these protocols crucially depends on the switch state, adding qubit relaxation processes prevents a simple expression for the fidelity, contrary to the case of the standard quantum state transfer (cf. Eq.~\eqref{eq:Fqst}). For this reason, we employ a quantum trajectory approach~\cite{Dalibard1992,Plenio1998} to numerically simulate the impact of a finite $T_1$ time, while photon loss is added on the averaged quantum state $\rho$. For each set of parameters we perform $N=1000$ quantum trajectories, and then bootstrapping with $R=100$ repetitions of $M=500$ randomly chosen samples to estimate the average and the standard deviation of the fidelity with respect to the target entangled state. 

We start by considering a Bell state between qubits 1 and 2 of the network, while the rest of the elements are idle. The quantum switch at the emitter node 1 is open, $\ket{1}_{s_1}$, with a dispersive shift $\chi_{s_1}=\kappa$, thus allowing only for half of the initial excitation in qubit 1 to be transferred (cf. Sec.~\ref{ss:Bell}). The receiver switch must be closed so that the incoming photon is fully absorbed, $\ket{0}_{s_2}$. For each of the $R$ repetitions, we obtain the average state $\rho^{(r)}$ over a set of randomly chosen samples among the $N$ trajectories, denoted $\mathcal{S}^{(r)}$, so that $\rho^{(r)}=\frac{1}{M}\sum_{j\in \mathcal{S}^{(r)}}|\psi_j(\tau/2)\rangle\langle\psi_j(\tau/2)|$.  For each repetition $r$ we obtain the Bell state fidelity as $F_{Bell}^{(r)}=\max_\theta \langle \Psi^+(\theta)|\rho_{1,2}^{(r)} |\Psi^+(\theta)\rangle$ being $\rho_{1,2}^{(r)}$ the reduced state of the first and second node qubits, including already photon loss, and $\ket{\Psi^+(\theta)}=\frac{1}{\sqrt{2}}(\ket{01}+e^{i\theta}\ket{10})$ is the targeted Bell state with a relative phase to account for dynamical phases during the protocol. Photon loss is introduced as a standard amplitude damping channel with probability $p_{loss}$. In this particular case, only the second qubit in an excited state is affected when a photon is lost. The final fidelity $F_{Bell}$ is obtained as the average over the $R$ repetitions, $F_{Bell}=\frac{1}{R}\sum_{r=1}^RF_{Bell}^{(r)}$, and the error is estimated using the standard deviation. The results are plotted in Fig.~\ref{fig3}(a) as a function of the $T_1$ noise affecting equally all the qubits in the network. The results show that already for conservative relaxation times $T_1\approx 100\ \mu$s the fidelity is above $99\%$, while photon loss $p_{loss}$ appears as the main limiting factor for $T_1\gtrsim 10^3\ \mu$s as the infidelity saturates to $1-F_{Bell}\approx p_{loss}/2= 6\cdot 10^{-4}$. 

We now move our attention to the generation of multipartite entangled states, namely, the GHZ and W states. As discussed in Sec.~\ref{ss:multi}, a $|{\rm GHZ}_3\rangle$ state can be approximately prepared between two qubit nodes and the emitter quantum switch. Besides potential imperfections introduced by decoherence effects, the preparation of a GHZ is limited by the physical constraint on the largest achievable ratio $\chi/\kappa$. The theoretical expression for the fidelity is given in Eq.~\eqref{eq:FGHZ}. Proceeding in a similar way as for the Bell state, we compute the reduced state for the qubits 1, 2 and the emitter switch, $\rho_{1,2,s_1}^{(r)}$ upon an error-free local $X_2$ gate on the qubit 2, for the $r$th repetition, also adding photon loss. We compute the state fidelity as $F_{GHZ}^{(r)}=\max_\theta \langle {\rm GHZ}_3(\theta)|\rho_{1,2,s_1}^{(r)} |{\rm GHZ}_3(\theta)\rangle$ with $|{\rm GHZ}_3(\theta)\rangle=\frac{1}{\sqrt{2}}(\ket{000}+e^{i\theta}\ket{111})$. As before, $F_{GHZ}$ is obtained as the average over the $R$ repetitions, $F_{GHZ}=\frac{1}{R}\sum_{r=1}^RF_{GHZ}^{(r)}$, and the error is estimated using the standard deviation. Fig.~\ref{fig3}(b) shows the results of $1-F_{GHZ}$ as a function of $\chi/\kappa$ for three different illustrative $T_1$ times. For $\chi/\kappa=1$ and $T_1=10 \ \mu$s, the fidelity $F_{GHZ}\approx 0.7>1/2$ already witnesses GHZ-entanglement in the final state, while for $T_1\gtrsim 100 \ \mu$s the theoretical expression Eq.~\eqref{eq:FGHZ} is saturated, leading to $F_{GHZ}\gtrsim 0.99$ for values $\chi/\kappa\gtrsim 5$ of the ratio between the dispersive and decay rate. 

The final target state is a $|W_3\rangle$ among the three node qubits 1, 2 and 3. As explained in Sec.~\ref{ss:multi} we employ a sequential protocol, and therefore we set a total time to $\tau=2\tau_{\rm opt}$. In this manner, the first emission-absorption process between qubits 1 and 2 takes a time $\tau_{\rm opt}$, and similarly for the subsequent process between qubits 2 and 3.  We compute the reduced state for qubits 1, 2 and 3 for the $r$th repetition, $\rho_{1,2,3}^{(r)}$, adding photon loss and obtain the fidelity as $F_{W}^{(r)}=\max_{\theta_1,\theta_2}\langle W_3(\theta_1,\theta_2)|\rho_{1,2,3}^{(r)}|W_3(\theta_1,\theta_2)\rangle$ with $|W_3(\theta_1,\theta_2)\rangle=\frac{1}{\sqrt{3}}(\ket{100}+e^{i\theta_1}\ket{010}+e^{i\theta_2}\ket{100})$ to account for local phases accumulated during the protocol. The average fidelity is plotted in Fig.~\ref{fig3}(c) as a function of the $T_1$ noise for all qubits. Note that photon loss must be taken into account for the first and second emission-absorption processes, and therefore, it limits the fidelity to a value $1-F_W\approx p_{loss}$, larger than for a Bell state. As for Bell states, the fidelity is above $99\%$ for realistic coherence times $T_1\approx 100\ \mu$s, while photon-loss limited fidelity $F_W\approx 10^{-3}$ is saturated for coherence times $T_1\gtrsim 600\ \mu$s.

\section{Conclusions}\label{s:conc}
In this article we put forward the use of additional qubit registers dispersively coupled to transfer resonators in a superconducting-circuit implementation of a quantum network. As we show, these extra qubits act as quantum switches that condition the dynamical evolution of the network in a coherent manner, allowing for deterministic single-photon routing and engineering of entangled states among distant nodes. The quantum switch induces a frequency shift in the transfer resonator dependent on the state of the switch. A closed switch enables the full-excitation transfer from a node qubit into a propagating single photon, as required in standard state transfer protocols. To the contrary, an open switch allows only for a partial transmission of the excitation. 

Employing wavepacket shaping techniques, where the coupling between emitter qubit and transfer resonator can be tuned in time, we show that the propagating photon acquires the same shape regardless of the dispersive shift, and thus, the external pulses to emit and absorb the single photon are agnostic to the switch state. This enables entangling operations conditioned to the state of the switches in the network. Based on our analytical derivation for the transmission amplitudes, that dependent solely on the ratio between the dispersive shift and the decay rate of the resonator, we detail the requirements for the generation of Bell pairs, GHZ and W states among spatially separated qubits.

The suitability of the proposed protocols leveraging quantum switches is supported by means of detailed numerical simulations for a three-node linear quantum network, based on recent experimental results, see for example Ref.~\cite{Storz2023}. In particular, we consider nodes connected by $L=10$ m long waveguides operating a mK temperatures. We numerically simulate all the elements of the network and include both $T_1$ noise and photon loss across the quantum link. Our numerical results illustrate that already for reasonable coherence times, $T_1\gtrsim 100\ \mu$s, the fidelities of target Bell and W states are above $99\%$, while for longer coherence times, photon loss in microwave waveguides, that feature a typical attenuation of $0.5$ dB/km~\cite{Qiu2025,Storz2023}, appear as the main limitation. To the contrary, our protocol aimed at generating a three-qubit GHZ is constrained by the ratio between dispersive shift and decay rate. 

This work paves the way for promising advancements in deterministic and rapid protocols between distant nodes in quantum networks, leveraging quantum switches to condition in a fully-coherent manner entanglement distribution and single-photon routing. These quantum switch-based protocols might unlock new possibilities in distributed quantum computation and quantum communication tasks. 
 
\begin{acknowledgments}
 RP acknowledges the Ram{\'o}n y Cajal (RYC2023-044095-I) research fellowship. 
\end{acknowledgments}

\appendix

\section{Analytical solution for the emission protocol}\label{app:a}
Under the approximation of a Markovian decay of the transfer resonator into the quantum link at rate $\kappa$, the equations of motion for the emitter node are given in Eqs.~\eqref{eq:qt}-\eqref{eq:ct}. For the $q_s=0$ case, i.e. when the switch is closed so that emitter qubit and resonator are placed at the same frequency, the emission of a sech-like photon, $|\gamma(t)|=\sqrt{\kappa/4}{\rm sech}(\kappa t/2)$ with bandwidth $\kappa$ is realized with a real-valued control $g(t)=\kappa/2 \ {\rm sech}(\kappa t/2)$~\cite{Penas2022}. In the following lines we show that the injected photon when $q_s=1$ acquires the same shape as for $q_s=0$, and derive the expression for the transmission probability as a function of $\chi$ and $\kappa$. From Eqs.~\eqref{eq:qt}-\eqref{eq:ct}, and using $g(t)=\kappa/2 \ {\rm sech}(\kappa t/2)$, one can find the corresponding equation of motion for $q(t)$, which is given by
\begin{align}\label{eq:ddqt}
    \ddot{q}(t)=\frac{1}{4}&\left\{ -\kappa^2 {\rm sech}^2(\kappa t/2)q(t)\right.\\&\left.- 2(i\chi+\kappa+\kappa \tanh(\kappa t/2))\dot{q}(t)\right\}.
\end{align}
The previous second-order differential equation for the emitter amplitude can be solved employing the ansatz $q(t)=a \tanh(\kappa t/2)+b$ with $a,b\in\mathbb{C}$ complex coefficients. Plugging the ansatz into Eq.~\eqref{eq:ddqt} and imposing the initial conditions  $\lim_{t\rightarrow -\infty}q(t)=1$ and  $\lim_{t\rightarrow -\infty}\dot{q}(t)=0$, we obtain the solution
\begin{align}\label{eq:qtsol}
    q(t)=\frac{2\chi+i\kappa(\tanh(\kappa t/2)-1)}{2(\chi-i\kappa)}.
\end{align}
In this manner, the qubit amplitude at the end of the emission protocol becomes
\begin{align}
    \lim_{t\rightarrow \infty}q(t)=\frac{\chi}{\chi-i\kappa}.
\end{align}
The excitation probability that remains in the qubit is $p_r\equiv \lim_{t\rightarrow \infty}|q(t)|^2$, only dependent  on the ratio $\chi/\kappa$ as mentioned in the main text,
\begin{align}
    p_r=\frac{1}{1+(\kappa/\chi)^2}.
\end{align}
Note that $\chi=0$ corresponds to a similar scenario as having a closed switch since emitter and transfer resonator have the same frequency. As expected due to the starting assumption to derive the control $g(t)$,  in this case $p_r=0$ as all the excitation is injected in the form of a traveling photon.  To the contrary, for a highly-detuned resonator $\chi/\kappa\rightarrow \infty$, the protocol fails to emit the photon and one finds $p_r=1-\frac{\kappa^2}{\chi^2}+O\left(\frac{\kappa^4}{\chi^4}\right)$.

The solution for the resonator amplitude can then be easily worked out from $q(t)$. From Eqs.~\eqref{eq:qt} and~\eqref{eq:qtsol}, it follows 
\begin{align}
    c(t)=\frac{1}{2}\frac{\kappa{\rm sech}(\kappa t/2)}{i\kappa-\chi}.
\end{align}
Moreover, the input-output relation $\gamma(t)=-i\sqrt{\kappa}c(t)$ allows us to  find the photon shape, $\gamma_\chi(t)=\sqrt{\kappa/4}{\rm sech}(\kappa t/2)(i\kappa)/(\chi-i\kappa)$, or similarly
\begin{align}
    |\gamma_\chi(t)|^2=\frac{\kappa }{4}{\rm sech}^2(\kappa t/2) \frac{\kappa^2}{\chi^2+\kappa^2},
\end{align}
where the subscript $\chi$ has been added to stress its dependence on the dispersive frequency shift. 
This important fact indicates that the shape of the photon is preserved, as only the normalization prefactor is modified. This prefactor accounts for the transmission probability, $p_t=1-p_r=(\kappa/\chi)^2 (1+(\kappa/\chi)^2)^{-1}$. That is, we find
\begin{align}
    |\gamma_\chi(t)|^2=p_t|\gamma_{\chi=0}(t)|^2.
\end{align}
For $\chi=0$ one recovers the resonant case, i.e. $\gamma_{\chi=0}(t)=-\sqrt{\kappa/4}{\rm sech}(\kappa t/2)$, and therefore the coefficients $\alpha$ and $\beta$ introduced in Eq.~\eqref{eq:s2} read as 
\begin{align}
    \alpha=\frac{\chi}{\chi-i\kappa}, \ \ \beta=\frac{-i\kappa}{\chi-i\kappa},
\end{align}
that are given in the main text. 

Finally, let us stress that the same effect would not be achieved for a generic control $g(t)$. As an example, consider the case in which the targeted photon shape is $|\gamma(t)|=\sqrt{\kappa'/4}{\rm sech}(\kappa' t/2)$ with $\kappa'<\kappa$, i.e. a sech-like photon with a reduced bandwidth with respect to the decay rate of the resonator. The control in this case can be also worked out, which becomes $g(t)=\frac{\kappa-\kappa'\tanh(\kappa't/2)}{2\sqrt{(1+e^{-\kappa't)\kappa/\kappa'-1}}}$~\cite{Penas2022}. In this situation, the control remains open in the long time limit, $\lim_{t\rightarrow \infty}g(t)\neq 0$. As a consequence, the emission process exhausts the qubit excitation regardless of the dispersive frequency shift, thus rendering the presence of the switch irrelevant. Even if the control $g(t)$ is switched off after some time to ensure a remaining non-zero excitation in the emitter, $p_r\neq 0$, the shape of the generated photon largely deviates from the desired form when $q_s=1$. This dependency on $\chi$ makes the emitted photons to be distinguishable dependent on the state of the switch, invalidating the operation outlined in Eqs.~\eqref{eq:s1}-\eqref{eq:s2}.


%

\end{document}